\documentclass[runningheads,UKenglish]{llncs}
\usepackage[T1]{fontenc}
\usepackage[utf8]{inputenc}
\usepackage{graphicx}
\usepackage{xcolor}

\usepackage[main=UKenglish]{babel}
\usepackage{xspace}
\usepackage[%
  group-minimum-digits=4,%
  list-final-separator={, and },%
  print-zero-integer=true,%
  free-standing-units,%
  round-mode=figures,%
  round-precision=3,%
  mode=match,%
  reset-text-series=false,%
  text-series-to-math=true,%
  uncertainty-mode=separate,%
  round-pad=false,%
]{siunitx}
\usepackage[]{subcaption}
\usepackage{listings}
\lstdefinestyle{inlinelst}{%
  basicstyle=\normalsize\ttfamily,%
  keywordstyle=,%
  language=Python,%
}
\lstdefinestyle{numberedlst}{%
  language=Python,%
  basicstyle=\scriptsize\ttfamily,%
  numbers=left,%
  numberstyle=\tiny,%
  tabsize=2,%
  breaklines=true,%
  xleftmargin=1em,%
  xrightmargin=1em,%
}
\lstdefinestyle{nonnumberedlst}{%
  language=Python,%
  basicstyle=\scriptsize\ttfamily,%
  numbers=none,%
  tabsize=2,%
  breaklines=true,%
  xleftmargin=1em,%
  xrightmargin=1em,%
}
\lstdefinestyle{nonnumberedtxt}{%
  basicstyle=\scriptsize\ttfamily,%
  numbers=none,%
  tabsize=2,%
  breaklines=true,%
  xleftmargin=1em,%
  xrightmargin=1em,%
}
\NewDocumentCommand{\inlinelst}{m}{\lstinline[style=inlinelst]{#1}}
\usepackage{colortbl}
\usepackage{adjustbox}
\usepackage{multirow}
\usepackage{todonotes}
\usepackage{csquotes}
\usepackage{mismath}
\def\BibTeX{{\rm B\kern-.05em{\sc i\kern-.025em b}\kern-.08em
T\kern-.1667em\lower.7ex\hbox{E}\kern-.125emX}}
\usepackage{hyperref}
\usepackage{color}

\usepackage[capitalise]{cleveref}
\usepackage{makecell}
\usepackage{cite}
\usepackage{booktabs}

\usepackage{enumitem}
\newlist{rqlist}{enumerate}{1}
\setlist[rqlist,1]{label=\textbf{RQ\arabic*}, ref=RQ\arabic*, nosep, left=0pt}

\let\origfootnote\footnote
\renewcommand{\footnote}[1]{\kern.06em\origfootnote{#1}}
\newcommand{\punctfootnote}[1]{\kern-.20em\origfootnote{#1}}

\usepackage{tcolorbox}
\usepackage{xspace}
\usepackage{relsize}

\newcommand{\toolname}[1]{%
  \textsc{#1}\xspace%
}

\newcommand{\pynguin}{%
  \toolname{Pynguin}%
}

\newcommand{\Pynguin}{%
  \toolname{Pynguin}%
}

\newcommand{\tensorflow}{%
  \toolname{Ten\-sor\-Flow}%
}

\newcommand{\pyTorch}{%
  \toolname{PyTorch}%
}

\newcommand{\muTester}{%
  \toolname{MuTester}%
}

\newcommand{\MuTester}{%
  \toolname{MuTester}%
}

\newcommand{\pyEvosuite}{%
  \toolname{Py\-Evo\-suite}%
}

\newcommand{\PyEvosuite}{%
  \toolname{Py\-Evo\-suite}%
}

\newcommand{\pynguinml}{%
  \toolname{Pynguin\-ML}%
}

\newcommand{\Pynguinml}{%
  \toolname{Pynguin\-ML}%
}

\newcommand{\docTer}{%
  \toolname{DocTer}%
}

\newcommand{\randoop}{%
  \toolname{Randoop}%
}

\newcommand{\numpy}{%
  \toolname{NumPy}%
}

\newcommand{\mxnet}{%
  \toolname{MXNet}%
}

\newcommand{\SUT}{SUT}
\newcommand{\SUTs}{SUTs\xspace}

\newcommand\effectsize{\ensuremath{\hat{A}_{12}}\xspace}
\newcommand\cohens{\ensuremath{\kappa}\xspace}

\newcommand*{\eg}{e.g.\@\xspace}
\newcommand*{\ie}{i.e.\@\xspace}

\NewDocumentEnvironment{summary}{m}{%
  \begin{tcolorbox}[title={Summary~(#1)}]%
  }{%
  \end{tcolorbox}%
}

\newcommand{\exnum}[1]{\num[round-mode=none]{#1}}
\newcommand{\perc}[1]{\qty{#1}{\percent}}

\newcommand{\subprocessexec}{%
  sub\-proc\-ess-exe\-cu\-tion\xspace
}

\newcommand{\FFI}{FFI\xspace}

\newcommand\CPP{%
  C\nolinebreak[4]\hspace{-.05em}%
  \raisebox{.4ex}{\relsize{-3}{\textbf{++}}}\xspace%
}

\newcommand{\fault}[1]{\textit{#1}\xspace}

\newcommand{\segfaults}{\fault{segmentation faults}}

\newcommand{\valid}{compliant\xspace} %
\newcommand{\Valid}{Compliant\xspace}
\newcommand{\VALID}{Compliant\xspace}
\newcommand{\Val}{Comp.\xspace}
\newcommand{\invalid}{non-compliant\xspace} %

\newcommand{\InVal}{Non-Comp.\xspace}

\newcommand{\modules}{\exnum{165}\xspace}
\newcommand{\modulesTorch}{\exnum{53}\xspace}
\newcommand{\modulesTorchTotal}{\exnum{2134}\xspace}
\newcommand{\modulesTF}{\exnum{112}\xspace}
\newcommand{\modulesTFTotal}{\exnum{1277}\xspace}

\newcommand{\mutesterTorchCoverage}{32.97\xspace}
\newcommand{\mutesterTFCoverage}{29.12\xspace}
\newcommand{\mutesterCoverage}{31.4}
\newcommand{\mutesterPynguinTorchCoverage}{27.8}
\newcommand{\mutesterPynguinTFCoverage}{24.07}
\newcommand{\mutesterPynguinCoverage}{26.3}
\newcommand{\pynguinInvalidPerc}{\perc{0.2686567}}%
\newcommand{\pynguinmlInvalidPerc}{\perc{0.266447}}%

\newcommand{\MlCoverage}{38.570288}

\newcommand{\MlBetterSig}{\exnum{77}\xspace}
\newcommand{\MlBetter}{\exnum{108}\xspace}

\newcommand{\MlWorse}{\exnum{34}\xspace}
\newcommand{\MlWorseSig}{\exnum{9}\xspace}
\newcommand{\MlEffectSize}{\num{0.6753513962206078}\xspace}
\newcommand{\MlRelativeCoverage}{\qty{63.903196906582814}{\percent}\xspace}

\newcommand{\PyCoverageShort}{33.4}
\newcommand{\MlCoverageShort}{38.6}
\newcommand{\PyCoverageShortTorch}{36.093614}
\newcommand{\MlCoverageShortTorch}{40.826748}
\newcommand{\PyCoverageShortTensorflow}{30.757548}
\newcommand{\MlCoverageShortTensorflow}{36.313828}
\newcommand{\PynguinGenTf}{\exnum{154}\xspace}
\newcommand{\PynguinGenTorch}{\exnum{114}\xspace}

\newcommand{\PynguinInvalTf}{\exnum{42}\xspace}
\newcommand{\PynguinInvalTorch}{\exnum{30}\xspace}

\newcommand{\PynguinValidTf}{\exnum{107}\xspace}
\newcommand{\PynguinValidTorch}{\exnum{75}\xspace}

\newcommand{\PynguinmlGenTf}{\exnum{180}\xspace}
\newcommand{\PynguinmlGenTorch}{\exnum{124}\xspace}

\newcommand{\PynguinmlInvalTf}{\exnum{45}\xspace}
\newcommand{\PynguinmlInvalTorch}{\exnum{36}\xspace}

\newcommand{\PynguinmlValidTf}{\exnum{121}\xspace}
\newcommand{\PynguinmlValidTorch}{\exnum{80}\xspace}

\providecommand{\done}[1]{}

\begin{document}

\title{Constraint-Guided Unit Test Generation\\for Machine Learning Libraries}

\author{Lukas Krodinger\inst{1}\orcidID{0009-0005-4571-4757} \and
  Altin Hajdari\inst{1}\orcidID{0009-0007-7953-6244} \and
  Stephan Lukasczyk\inst{1,2}\orcidID{0000-0002-0092-3476} \and
Gordon Fraser\inst{1}\orcidID{0000-0002-4364-6595}}
\authorrunning{L. Krodinger et al.}

\institute{%
  University of Passau, Germany
  \and
  JetBrains Research, Germany
}

\maketitle

\begin{abstract}
  Machine learning (ML) libraries such as \pyTorch and \tensorflow
  are essential for a wide range of modern applications.
  Ensuring the correctness of ML libraries through testing
  is crucial.
  However, ML APIs often impose strict input constraints involving
  complex data structures such as tensors. Automated test generation
  tools such as \pynguin are not aware of these constraints and often
  create \invalid inputs.  This leads to early test failures and
  limited code coverage.
  Prior work has investigated extracting constraints from official API
  documentation.  In this paper, we present \pynguinml, an approach
  that improves the \pynguin test generator to leverage these
  constraints to generate \valid inputs for ML APIs, enabling more
  thorough testing and higher code coverage.
  Our evaluation is based on \modules modules from \pyTorch and
  \tensorflow, comparing \pynguinml against \pynguin.
  The results show that \pynguinml significantly improves test
  effectiveness, achieving up to \MlRelativeCoverage higher code
  coverage.

\end{abstract}

\section{Introduction}\label{sec:introduction}

Machine learning (ML) libraries such as
\pyTorch
\footnote{%
  \href{https://pytorch.org/}{\url{https://pytorch.org/}},
  last accessed 2025–07–29.%
}
and
\tensorflow,\punctfootnote{%
  \href{https://tensorflow.org/}{\url{https://tensorflow.org/}},
  last accessed 2025–07–29.%
}
are used in many real-world applications
and downloaded thousands of times each day.%
\punctfootnote{%
  \href{https://pypistats.org/packages/pytorch}{\url{https://pypistats.org/packages/pytorch}}
  and \newline
  \href{https://pypistats.org/packages/tensorflow}{\url{https://pypistats.org/packages/tensorflow}},
  last accessed 2025–07–29.%
}
To ensure that these libraries work as
intended, developers typically write unit tests. However, manually
creating effective unit tests is a tedious and time-consuming
task~\cite{runesonSurveyUnitTesting2006}.
Thus, automated unit test generation tools have been introduced.
Prominent examples of existing tools include \randoop~\cite{Pacheco2007} and
\toolname{EvoSuite}~\cite{fraser_evosuite_2011} for
Java or \pynguin~\cite{lukasczykPynguinAutomatedUnit2022} for Python.
These tools typically analyse a \emph{subject under test}~(\SUT), such as a Java
class or Python module, and use meta-heuristic search algorithms to
generate tests. The goal is to create inputs that trigger
the \SUT's routines (i.e., functions, methods, and constructors) in a
manner that maximizes a fitness metric, most commonly branch coverage.

However, many ML libraries require specific input constraints,
which current test generation tools do not
consider~\cite{wangAutomaticUnitTest2021, zhangMachineLearningTesting2022}.
Inputs must follow a particular structure, such as a \inlinelst{list},
\mbox{\inlinelst{tuple}}, or \inlinelst{tensor} and they must satisfy
certain properties like a specific data type or shape.  The most
popular language for ML applications is Python, yet Python's
state-of-the-art test generation tool \pynguin is not aware of these
constraints.  Thus, many generated test cases include API calls with
\invalid inputs, which often trigger early failures. This results in
low code coverage. For example, consider \cref{fig:glu_code}, which
shows a simplified version of the PyTorch API
\inlinelst{torch.nn.functional.glu}, which calls the \inlinelst{dim()}
method on the parameter \inlinelst{input}, assuming it is a valid
tensor. An example test case generated by \pynguin (\cref{fig:glu_pynguin})
calls this function with \inlinelst{None} as input, which is not a
valid tensor and causes an exception to be raised.  As \pynguin
generates test suites where all test cases pass by construction, it
marks this test with \inlinelst{@mark.xfail(strict=True)} from
\toolname{PyTest}, which results in the test case to expect the exception and
thus pass. However, the test does not reach the core logic of the function and
contributes little to code coverage.

\begin{figure}[t]
  \centering
  \begin{minipage}[t]{0.48\textwidth}
    \centering
\begin{lstlisting}[
      style=nonnumberedlst,%
      frame=single,%
    ]
def glu(
    input: Tensor, dim: int = -1
) -> Tensor:
    if input.dim() == 0:
        raise RuntimeError()
    # Logic of the function ...
\end{lstlisting}
    \subcaption{Simplified code of the PyTorch API
    \texttt{torch.nn.functional.glu}.}
    \label{fig:glu_code}
  \end{minipage}
  \hfill
  \begin{minipage}[t]{0.48\textwidth}
    \centering
\begin{lstlisting}[
      style=nonnumberedlst,%
      frame=single,%
    ]
import torch.nn.functional as F

@mark.xfail(strict=True)
def test_case_0():
    none_type_0 = None
    F.glu(none_type_0)
\end{lstlisting}
    \subcaption{Typical test generated by \pynguin with
    \invalid input parameters.}
    \label{fig:glu_pynguin}
  \end{minipage}
  \caption{Example of a failing test case due to missing input constraints.}
  \label{fig:glu_invalid_example}
\end{figure}

\begin{figure}[!t]
  \centering
\begin{lstlisting}[
    style=nonnumberedlst,%
    frame=single,%
  ]
import numpy
import torch
import torch.nn.functional as F

def test_case_0():
    var_0 = [
        [15.06, 1038.1, 1695.079, -97.7],
        [12.06, 0.1179, 16.6, 32.23]
    ]
    var_1 = "float32"
    var_2 = numpy.array(object=var_0, dtype=var_1)
    var_3 = torch.tensor(data=var_2)
    F.glu(var_3)
\end{lstlisting}
  \caption{\label{fig:glu_valid_test}%
    \Valid test generated by \pynguinml using API constraints.
  }
\end{figure}

To address this problem, we leverage API constraints
previously extracted from the
documentation~\cite{xieDocTerDocumentationguidedFuzzing2022} of libraries
to guide test generation.
These constraints specify the
expected structure and properties of input arguments, such as \textit{data
type}, tensor \textit{shape}, or \textit{dimension} values. By
incorporating this information into the test generation process,
we can create \valid inputs that satisfy these constraints. This enables deeper
parts of the API code to be reached during testing.
We implement our approach, \pynguinml, as an extension of \pynguin.

\Cref{fig:glu_valid_test} shows a test case generated by \pynguinml
for the \inlinelst{glu} function from \cref{fig:glu_code}.  The API
constraints for the \inlinelst{glu} function tell \Pynguinml that a
tensor is required as well as its expected \textit{dimension},
\textit{shape}, and \textit{data type}. \Pynguinml generates a nested
list that adheres to these structural properties and assigns the
correct \numpy data type as a string. It then converts the list to a
\numpy array and finally to a \pyTorch tensor. The generated input
satisfies the API's requirements, thus executing the \SUT\xspace
beyond input validation.

To evaluate our approach, we identified 165 modules from \pyTorch and
\tensorflow for which API constraint data is available. We compare
\pynguinml to \pynguin and to the existing
\muTester~\cite{narayananAutomaticUnitTest2023a}
tool that also uses
API constraints. Our evaluation focuses on two key metrics: code
coverage and the number of \valid test cases, i.e., test cases where
input parameters do not violate the documented API constraints.
We make the following contributions:
\begin{itemize}
  \item We propose \pynguinml, an extension of \pynguin that
    incorporates constraints from the documentation of libraries
    into the test generation process.
  \item We introduce new statement types tailored to ML-specific
    needs, such as nested lists with valid shapes, enum values, and
    unsigned integers.
  \item We evaluate \pynguinml compared to \pynguin and \muTester
    on 165 modules from \pyTorch and \tensorflow.
\end{itemize}

Our evaluation shows that \pynguinml significantly improves
test generation for Python ML libraries compared to \pynguin and \muTester.

\section{Background}\label{sec:background}
Our work is based on unit test generation for Python
and is tailored to libraries with extensive API documentation,
specifically machine learning~(ML) libraries.

\subsection{Machine Learning Libraries in Python}

Machine Learning (ML) is the foundation for many applications,
such as image classification, natural language processing, and
recommendation systems.
To facilitate ML development open-source libraries like \tensorflow and
\pyTorch provide standardized APIs that simplify complex algorithms.
To ensure correct usage of these APIs, ML libraries provide detailed
documentation.
A key part of the API documentation defines constraints for input parameters.
The two main categories are: (1) the required data structures and (2) the
properties of these data structures.
Firstly, parameters must follow specific data structures, which includes
built-in types such as \inlinelst{int}, \inlinelst{str}, or \inlinelst{list}
and library-specific types like \tensorflow's
\inlinelst{tf.Tensor}\footnote{%
  \href{https://tensorflow.org/api_docs/python/tf/Tensor}{\url{https://tensorflow.org/api_docs/python/tf/Tensor}},
  last accessed 2025–08–21.%
}
or \pyTorch's \inlinelst{torch.Tensor}.\punctfootnote{%
  \href{https://pytorch.org/docs/stable/tensors.html}{\url{https://pytorch.org/docs/stable/tensors.html}},
  last accessed 2025–08–21.%
}
If the input does not match the expected type, the API function typically
raises an error during input validation.
Secondly, these data structures must satisfy several specific
properties, so-called constraints. One property of the constraints is the
\textit{data type}, like \inlinelst{float} or the library-specific
\inlinelst{torch.float32}.
Another property is the \textit{shape}, which defines the dimensions
as a tuple of integers indicating the size along each axis.
A tuple's length corresponds to the \textit{dimension}
of the tensor; \eg, a shape of \inlinelst{(3, 4)} describes a
2-dimensional tensor.
Furthermore, a valid value \textit{range} defines the
allowable set of values for a parameter, \eg, restricting a \inlinelst{float}
to the interval \inlinelst{[0, 1]} or a string to a predefined list of
supported values.

While these ML constraints are crucial for correct API usage and
essential for automated test generation,
they are often not explicitly defined in the code but rather
documented informally using different formats
and conventions across libraries.
The constraints must be extracted and normalized into a consistent,
standardized form. This poses several challenges that were addressed
in prior work~\cite{xieDocTerDocumentationguidedFuzzing2022,
shiACETestAutomatedConstraint2023, narayananAutomaticUnitTest2023a}.
We use previously extracted
constraints~\cite{xieDocTerDocumentationguidedFuzzing2022} for the ML
libraries \tensorflow and \pyTorch
for evaluating our approach.

\subsection{Unit Test Generation for Python}

Creating effective test suites is critical for reducing the
risk of software failures. However, manually creating test
cases is often time-consuming and labour-intensive. To alleviate this
burden, numerous automated techniques for unit test generation have
been proposed~\cite{FR19}. Some approaches use random test
generation, sometimes enhanced by feedback from previous executions
to inform subsequent test creation~\cite{PLE+07}. Besides that,
search-based software engineering~(SBSE) methods, such as
evolutionary algorithms, have been employed to optimize test cases
with respect to a specified fitness function~\cite{Ton04}, such as
branch coverage.
For an in-depth discussion of search-based test generation and the underlying
algorithms, we refer to the extensive literature on this
topic~\cite{CGA18, Panichella2018}.

\Pynguin~\cite{lukasczykPynguinAutomatedUnit2022} is a state-of-the-art
test-generation framework designed for the Python programming language.
It incorporates multiple SBSE test-generation strategies,
including feedback-directed random test generation~\cite{PLE+07} and the
DynaMOSA~\cite{PKT18b} evolutionary algorithm.
\Pynguin aims to generate test cases that yield high coverage
values. To achieve this, it analyzes the \SUT\xspace and collects all public
classes, functions, and methods and saves all their lines or
branches~(depending on the configuration) as targets for the test
generation. Upon creating test cases for a focal method that has at least
one parameter, \pynguin needs to generate arguments to call the method with.
It randomly generates objects for built-in types, collections, and
classes defined in the \SUT\@.
For assembling objects of user-defined classes, \pynguin tries using
constructors or factory methods. If the selected constructor requires
more than one parameter, \pynguin repeats the process recursively~(up to a
pre-defined maximum recursion depth) until all parameters are satisfied.
During this process, \pynguin does not consider API constraints
defined in the API
documentation which leads to \invalid inputs.
In turn, the generated tests execute
less of the \SUT's code, which reduces the effectiveness of the generated tests.

\subsection{MuTester}
\MuTester~\cite{narayananAutomaticUnitTest2023a} is a test generation tool
based on \pynguin that, similarly to our approach, enhances test generation
for ML libraries.
\MuTester uses signature definition constraints mined from the method
signature using a regular expression based string analyzer, API
constraints mined from API documentation, and concrete input values mined
from example code snippets.
During test generation, \muTester iteratively matches candidate methods
to sequence patterns by checking for usage patterns based on the API
constraints.
It further searches past executions for matching inputs or
generates new ones based on the mined constraints if none are found.
In contrast to \MuTester, \pynguinml does not use
mined constraints from signatures. Type hints in the source code,
which are used by \pynguin if available, already provide the same information.
Furthermore, \pynguinml does not use mined constraints from example code
snippets, but instead focuses on constraints mined from API documentation.

\section{Test Generation Using API Constraints}\label{sec:approach}

To address the limitations of test generation for libraries with API
constraints---especially Machine Learning (ML) libraries---we extend
\pynguin~\cite{lukasczykPynguinAutomatedUnit2022}
by integrating API constraints into input generation, which we call
\pynguinml.
Before generating tests, \pynguinml analyzes the module under test
to identify used APIs based on the module and API name and loads the
corresponding constraints.
Each constraint file describes input requirements for a specific API.
These files specify properties per parameter, including:
\begin{itemize}
  \item \textbf{dtype}: the expected data type (e.g., \texttt{torch.float32}),
  \item \textbf{ndim}: required number of dimensions (rank),
  \item \textbf{tensor\_t}: whether the input must be a tensor,
  \item \textbf{structure}: required container types (e.g.,
    \texttt{list} or \texttt{tuple}),
  \item \textbf{range}: valid intervals for numeric parameters,
  \item \textbf{enum}: allowed values for categorical parameters.
\end{itemize}

When loading the constraints in the initial setup phase, a validation
process removes invalid ones.
This includes syntactically incorrect entries, undefined
types, or constraints that are logically contradictory, such as
specifying a negative integer as \textit{dimension} or an invalid numeric
\textit{range}, \eg, \inlinelst{[1, 0]}.

The constraints guide the input generation of \Pynguinml to generate
\valid inputs and sequences. For primitives, \eg \inlinelst{float},
\pynguinml extends existing \pynguin mechanisms by applying
constraints such as \inlinelst{range}. For example, a
\inlinelst{float} parameter with a \inlinelst{range} of
\inlinelst{[0.0, 1.0]} results in a random \inlinelst{float} within
this \inlinelst{range}.
To support further constraints, \pynguinml introduces additional
statements: (1) an \textit{enum} statement for parameters with
\inlinelst{enum} values, selecting from the predefined set of allowed
values; (2) an \textit{unsigned-integer} statement for \inlinelst{dtype}
constraints that specify unsigned integer types, generating
non-negative integers; and (3) a \textit{nested-list} statement for
\inlinelst{dtype} constraints that specify tensors. This \textit{nested-list}
serves as a library-independent representation of tensors,
which \pynguinml uses in the following pre-defined sequence of statements
tailored towards ML libraries to generate a \valid tensor:
\begin{enumerate}
  \item Generate a \textit{nested-list} with a shape, as defined by
    the \inlinelst{ndim} constraint.
  \item Create a string statement storing the desired \numpy
    \inlinelst{dtype}.%
  \item Construct a \inlinelst{numpy.ndarray} from the list and
    \inlinelst{dtype}.
  \item Convert the array into a library-specific tensor using a
    user-defined function (e.g., \inlinelst{torch.tensor} or
    \inlinelst{torch.from\_numpy}).
\end{enumerate}

The new statements require modifications to the
evolutionary algorithm used by \pynguinml.
To prevent invalid test cases, we do not allow crossover operations
within tensor-generation sequences. If a crossover point occurs inside such a
sequence, \pynguinml discards the subsequent statements of the sequence
and replaces them with a newly generated sequence.
We added mutation functionality for the new
statements: a \textit{nested-list} mutates its elements while preserving its
\textit{shape}, an \textit{enum} selects a different allowed value, and an
\textit{unsigned integer} always mutates to a new non-negative value.
If a mutation deletes a tensor creation statement,
any subsequent dependent statements are also deleted.

While these modifications help producing \valid parameters when calling
functions, achieving high code coverage also requires \invalid inputs.
For example, input validation code that performs checks on the
parameters passed into a function can only be properly tested when
also allowing for \invalid parameter values (i.e., negative
tests). Consequently, when generating parameters for a function for
which API constraints are available, \pynguinml makes a probabilistic
choice of whether to produce a \valid parameter, or whether to fall
back to \pynguin's standard object generation mechanisms, which will
likely lead to \invalid parameters suitable for testing input
validation code.

\section{Evaluation}\label{sec:eval}

We extended \pynguin with support for API constraints
as described in \cref{sec:approach}. To evaluate the effectiveness
of our extension, \pynguinml, we ask:

\vspace{0.5em}
\begin{rqlist}
\item \label{rq:coverage} \textbf{Branch Coverage:}
  \emph{To what extent does \pynguinml
  improve code coverage compared to \pynguin and \muTester?}
\end{rqlist}

\vspace{0.5em}

We conjecture that the API constraints lead to more \valid inputs.
We define a test case as \emph{\valid} if it does not include
an API call with inputs that violate the constraints specified in the
API documentation.  To evaluate the impact of more \valid test cases,
we ask:

\vspace{0.5em}
\begin{rqlist}[resume]
\item \label{rq:invalid} \textbf{\VALID Tests:}
  \emph{How many more \valid test cases does \pynguinml
  generate compared to \pynguin?}
\end{rqlist}

\subsection{Evaluation Setup} \label{sec:eval:setup}
\subsubsection{Subjects} \label{sec:eval:subjects}
We evaluate our approach on ML libraries for which
constraint data is available~\cite{xieDocTerDocumentationguidedFuzzing2022},
specifically \pyTorch and \tensorflow.
The provided constraints correspond to \tensorflow v2.1.0 and \pyTorch
v1.5.0. However, we use \tensorflow v2.8.0 and \pyTorch v1.13.0 in our
evaluation, as these versions are compatible with Python 3.10, which
is required by \pynguin. Although
\docTer~\cite{xieDocTerDocumentationguidedFuzzing2022}
provides constraints for \mxnet as well, we exclude \mxnet from our
evaluation because
it does not support Python 3.10 and is retired.
Among all importable, non-test and non-\texttt{\_\_init\_\_.py}
modules~(\modulesTorchTotal for \pyTorch and \modulesTFTotal for
\tensorflow), we selected those modules for which we have a
constraint for at least one API
within the module, resulting in \modulesTorch modules from \pyTorch and
\modulesTF modules from \tensorflow, for a total of \modules modules.

\subsubsection{Experiment Settings} \label{sec:eval:settings}

\pyTorch and \tensorflow use Python's \emph{foreign function
interface}~(\FFI) to call low-level functions in C or \CPP for
performance reasons.  These functions can lead to low-level errors,
such as \segfaults, which cannot be caught by Python's exception
handling and instead cause the Python interpreter to crash, which in
turn terminates \pynguin.
We therefore configure \pynguin to execute the test cases in separate
subprocesses rather than only in separate threads.
This comes with the drawback of a performance
overhead.  To compensate for this, we set the \textit{search time} to
\qty{900}{\second} rather than previously used
\qty{600}{\second}~\cite{lukasczyk_automated_2020,
  lukasczykPynguinAutomatedUnit2022,
  lukasczykEmpiricalStudyAutomated2023, lemieux_codamosa_2023,
yang_llm-enhanced_2025}.  Additionally, we set the newly introduced
\textit{maximum tensor dimension} and the \textit{maximum size per
shape dimension} to \exnum{5}. This is sufficient for most ML APIs
and helps balance memory usage. All other configuration options use
the default settings of \pynguin.  We execute both \pynguinml and
\pynguin on each subject \exnum{20} times.
We use a \toolname{Docker} container to
isolate the executions from their environment based on Python~3.10.16
and execute all runs on dedicated servers, each equipped with an AMD EPYC 7443P
CPU and \qty{256}{\giga\byte} RAM. However, we only assign a single
CPU core and \qty{8}{\giga\byte} of RAM to each run to simulate an
environment closer to real-world operation conditions.

\subsubsection{Experiment Procedure}\label{sec:eval:procedure}
As a basis for answering all research questions, we execute both
\pynguinml and \pynguin on all 165 subject modules.
For every successful run, we collect the achieved branch coverage and
the generated test cases. We use the collected coverage data to answer
\ref{rq:coverage} by comparing between \pynguin and \pynguinml.  We
also compare our findings with those from \muTester, but as their
replication package is incomplete, we were unable to re-run \muTester
ourselves. We therefore compare the numbers reported in the \muTester
paper with a re-run of \pynguinml for \qty{300}{\second}~(as was done
for \muTester) for a fair comparison.
To address \ref{rq:invalid} we inspect a subset of the generated tests
by first randomly sampling \exnum{10} modules per ML library, and then
randomly selecting one test file (out of \exnum{20} repetitions) per
tool for each of these modules. All test cases violating documented
API constraints are then manually classified individually by two of
the authors of the paper.
Further details on our setup and analysis scripts can be found in our
replication package~\cite{dataset}.

\subsubsection{Evaluation Metrics} \label{sec:eval:metrics}

To assess by how much \pynguinml improves over \pynguin%
, we compute the relative coverage~\cite{AF13} as a
percentage.
Given the coverage of a subject~\(s\) in an execution~\(e\), denoted
by~\(\cov(s, e)\), we refer to the best coverage on~\(s\) in any execution
by~\(\max \left( \cov(s) \right)\) and to the worst coverage on~\(s\) in any
execution by~\(\min \left( \cov(s) \right) \).  We define relative
coverage~\( \cov_r(s, e) \) as
\[
  \cov_r(s, e) =
  \frac{%
    \cov(s, e) - \min \left( \cov(s) \right)%
  }{%
    \max \left( \cov(s) \right) - \min \left( \cov(s) \right)%
  }.\label{eq:background-relcov}
\]
In case \( \min \left( \cov(s) \right) = \max \left( \cov(s) \right) \),
i.e., minimum and maximum coverage for a subject~\(s\) are equal, we define
the relative coverage~\( \cov_r(s, e) = \qty{100}{\percent}\)~\cite{CGA18}.

To evaluate if one configuration performs better than another
configuration, we use the Mann-Whitney
U-test~\cite{mann_test_1947} with $\alpha = 0.05$. Additionally, we
compute the Vargha and Delaney effect size \effectsize~\cite{VD00} to
investigate the difference in the achieved overall coverage between
two configurations.  These metrics do not make assumptions on the data
distribution, are in line with recommendations for assessing
randomized algorithms~\cite{AB14}, and were used in previous
work~\cite{lukasczyk_automated_2020,
  lukasczykPynguinAutomatedUnit2022,
  lukasczykEmpiricalStudyAutomated2023, lemieux_codamosa_2023,
yang_llm-enhanced_2025}.

We analyse whether a test case is \valid per module, with
Cohen's Kappa coefficient~\cohens measuring inter-rater agreement in
manual validation~\cite{cohen1960coefficient}.

\subsection{Threats to Validity}
\subsubsection{Internal Validity}
There are several threats regarding the comparison with \muTester:
First, we use a newer version of \pynguin than the one used in the original
\muTester study. Our version may include performance improvements that
impact coverage and test generation behaviour. We did this to allow for
the usage of \subprocessexec.
Second, the usage of \subprocessexec allows isolating crashes during
test execution.
While this is necessary for stability when testing ML libraries, it introduces
additional overhead. However, when not using it the results for \pynguin
and \pynguinml can only be better due to omitting the performance overhead.
Third, the exact configuration, module selection, and constraints used by
\muTester are not publicly available, and thus the experiments might
have been executed under different conditions. We did everything possible
to make comparability as good as possible.

Besides the comparison with \muTester other threats are regarding the
manual analysis of \valid test cases.
The manual analysis introduces a risk of human error, as determining
whether inputs violate constraints requires interpreting API documentation.
To mitigate this, we determine the violations two times
independently, and only consider those as \valid where both agree.

\subsubsection{External Validity}

Our evaluation focuses on two widely used ML libraries: \pyTorch and
\tensorflow. While these libraries differ in implementation, they
share a similar structure and design commonly found in many ML
frameworks. Therefore, we expect \pynguinml to generalize
well to other ML libraries with similar characteristics. However, the
results may not generalize to libraries outside the ML domain, where
APIs and usage patterns differ considerably. Additionally, our
evaluation is limited to modules for which constraints were available,
because the replication package for \docTer does not include
the code for HTML scraping~\cite{xieDocTerDocumentationguidedFuzzing2022}.
Therefore, the results may not generalize to ML
libraries or APIs where constraint information is incomplete or less
accurate.

Furthermore, our evaluation of \valid test cases is based on
a limited sample of 10 randomly selected modules per library, which
may not fully capture the diversity of APIs in \pyTorch and
\tensorflow.

\subsubsection{Construct Validity}

Our approach relies on API constraints extracted by prior
work~\cite{xieDocTerDocumentationguidedFuzzing2022}, which
may not always be fully accurate or up to date.
The constraint versions differ from the versions we use for test
generation~(see \cref{sec:eval:subjects}).
While some APIs may have changed, the APIs are generally stable
enough for relatively small version gaps.
Furthermore, the constraints were extracted with an accuracy of
approximately \perc{85.4}~\cite{xieDocTerDocumentationguidedFuzzing2022},
\ie, some rules may be incorrect or missing.
This can lead to false assumptions such as incorrectly
rejecting \valid inputs or accepting \invalid ones. Despite these
limitations, we chose these constraints as they are publicly
available and were used in prior
work~\cite{xieDocTerDocumentationguidedFuzzing2022}.

\subsection{RQ1: Branch Coverage}

\begin{table}[t]
  \centering
  \caption{\label{tab:coverage-significance}%
    Number of modules where \pynguinml (treatment)
    performed (significantly) better, equal or (significantly) worse
    than \pynguin (control) and the Vargha and Delaney
    \effectsize~effect size.
  }
  \begin{tabular}{@{}lrrrS@{}}
  \toprule
  \textbf{Library} & \textbf{Better (Sig)} & \textbf{Equal} &
  \textbf{Worse (Sig)} & \textbf{\effectsize} \\
  \midrule
  \pyTorch & 30 (19) & 6 & 15 (3) & 0.630507 \\
  \tensorflow & 78 (58) & 11 & 19 (6) & 0.696528 \\
  \midrule
  \textbf{Total} & \textbf{108 (77)} & \textbf{17} & \textbf{34 (9)}
  & \textbf{0.675} \\
  \bottomrule
\end{tabular}

\end{table}

\begin{figure}[t]
  \centering
  \begin{subfigure}[t]{0.49\textwidth}
    \centering
    \includegraphics[width=\linewidth]{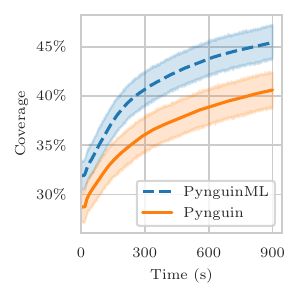}
    \caption{\pyTorch}
  \end{subfigure}
  \hfill
  \begin{subfigure}[t]{0.49\textwidth}
    \centering
    \includegraphics[width=\linewidth]{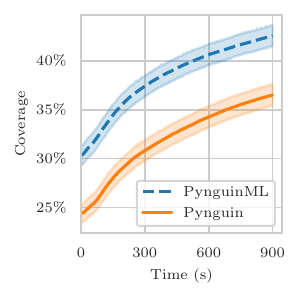}
    \caption{\tensorflow}
  \end{subfigure}

  \caption{Development of the coverage over time for \pynguin
  and \pynguinml.}
  \label{fig:coverage_over_time}
\end{figure}

To answer \ref{rq:coverage}, we compare
the development of coverage over time~(\cref{fig:coverage_over_time}):
\Pynguinml achieves higher branch coverage on both ML libraries than \pynguin.
The relative coverage of \pynguinml overall is \MlRelativeCoverage.
For both ML libraries, \pynguinml starts with a higher initial coverage
and maintains a clear lead throughout the generation
process~(\cref{fig:coverage_over_time}).
On \pyTorch, \pynguinml gains coverage more quickly in the early
phase, while on \tensorflow, it keeps a more consistent advantage from
\pynguin for the entire search time. This shows that
\pynguinml not only starts efficiently, but consistently
outperforms \pynguin.
These observations are also confirmed statistically as shown in
\cref{tab:coverage-significance}.  On \MlBetter~(\MlBetterSig
significant) modules, \pynguinml achieves a (significantly) higher
branch coverage than \pynguin, while on \MlWorse~(\MlWorseSig
significant) modules \pynguin achieves a (significantly) higher branch
coverage than with \pynguinml.  The Vargha and Delaney~\effectsize
effect size confirms a mean positive effect of \MlEffectSize.

\begin{figure}[t]
  \centering
  \begin{subfigure}[t]{0.49\textwidth}
    \centering
    \includegraphics[width=\linewidth]{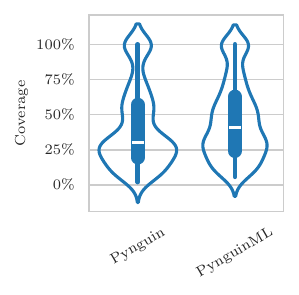}
    \caption{\pyTorch}
  \end{subfigure}
  \hfill
  \begin{subfigure}[t]{0.49\textwidth}
    \centering
    \includegraphics[width=\linewidth]{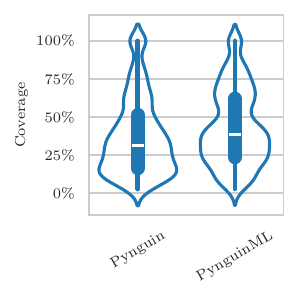}
    \caption{\tensorflow}
  \end{subfigure}
  \caption{Coverage distributions for \pynguin and \pynguinml.}
  \label{fig:coverage_distribution}
\end{figure}

While for average coverage (over time) \pynguinml outperforms
\pynguin, we want to further analyse how the coverage distribution
differs across different modules. \Cref{fig:coverage_distribution}
depicts coverage values over the full range of \perc{0} to \perc{100}.
We observe that \pynguinml tends to achieve higher and more
concentrated coverage values. Its distribution is denser around the
mid-to-high range, suggesting more consistent performance across
modules. In contrast, \pynguin shows a wider spread,
particularly towards lower coverage values and most notably for
\tensorflow. Overall, \pynguinml not only increases average
coverage compared to \pynguin, but it produces better
results across different modules.

\begin{table}[t]
  \centering
  \caption{Mean branch coverage of \pynguin, \muTester, \pyEvosuite and
    \pynguinml on \pyTorch and \tensorflow in \%. We weighted average values
    by the number of modules/code examples per library.
  }
  \centering
  \label{tab:mutester_coverage}
  \begin{tabular}{@{}lSSSS@{}}
    \toprule
    \textbf{Library} & \textbf{\Pynguin} & \textbf{\Pynguinml}
    & \textbf{\PyEvosuite} & \textbf{\MuTester} \\
    \midrule
    \pyTorch    & \PyCoverageShortTorch & \MlCoverageShortTorch &
    \mutesterPynguinTorchCoverage & \mutesterTorchCoverage \\
    \tensorflow & \PyCoverageShortTensorflow & \MlCoverageShortTensorflow &
    \mutesterPynguinTFCoverage & \mutesterTFCoverage \\
    \midrule
    \textbf{Average} & \textbf{\PyCoverageShort} & \textbf{\MlCoverageShort} &
    \textbf{\mutesterPynguinCoverage} & \textbf{\mutesterCoverage} \\
    \bottomrule
  \end{tabular}
\end{table}

We are further interested in how \pynguinml performs compared
to other state-of-the-art tools. Since
\muTester~\cite{narayananAutomaticUnitTest2023a} also employs a
constraint-based strategy to improve test generation for ML libraries,
we compare \pynguinml with \muTester.
\Cref{tab:mutester_coverage} shows the mean branch coverage
achieved by \pynguin, \muTester, \pyEvosuite, and \pynguinml
across both ML libraries. \PyEvosuite refers to the baseline
\pynguin used in the \muTester study~\cite{narayananAutomaticUnitTest2023a},
which in fact is a version of \pynguin.
\Pynguin achieves \perc{\PyCoverageShort},
while \pynguinml achieves \perc{\MlCoverageShort} branch coverage
on average across both libraries.
Despite not using \subprocessexec, which avoids a
performance overhead, \MuTester achieves \perc{\mutesterCoverage}
on average, which is lower than \pynguinml.
\pyEvosuite achieves \perc{\mutesterPynguinCoverage} on average,
which is lower than \pynguin.

\begin{summary}{\hypersetup{linkcolor=white}\ref{rq:coverage}:
  Branch Coverage}
  In our experiments, \pynguinml achieves higher branch coverage than
  \pynguin and \muTester.
\end{summary}

\paragraph*{Discussion}
Our results show that incorporating API constraints extracted from
the documentation of ML libraries significantly improves branch
coverage compared to unconstrained test generation. Test cases
generated by \pynguinml are more likely to pass internal API
validation checks, which enables them to reach code branches that
{\pynguin}'s test cases often miss.
We conclude that the increased coverage is due to the tensor-aware
\pynguinml generation loop that leverages the API constraints.

Despite these improvements, the achieved branch coverage remains
modest, with \pynguinml reaching on average
\perc{\MlCoverage}.
\Cref{fig:coverage_over_time} shows that the search has not converged
after \qty{900}{\second} which suggests that running the search
longer achieves higher coverage.
This is particularly relevant for ML libraries, where many
modules are large and internally complex. For example, the
\inlinelst{torch.nn.functional.interpolate} API has a cyclomatic
complexity of \exnum{38}, which, according to McCabe's
metric~\cite{mccabeComplexityMeasure1976},
indicates high structural complexity and thus greater testing difficulty.
Moreover, the \inlinelst{torch.nn.functional} module alone comprises
more than \exnum{80} APIs and over \exnum{5000} lines of code,
emphasizing the challenges of covering all branches
even when using API constraints. We conclude that the limited
improvement in coverage may be due to the complexity of the modules
under test, which might require even more sophisticated input generation
strategies to achieve more substantial coverage gains.

Another limiting factor is the required use of \subprocessexec, which
introduces considerable execution overhead. Testing such complex
modules effectively would likely require substantially more than
\qty{900}{\second} of generation time, but longer time budgets are
impractical in real-world development workflows. Future work should
therefore investigate more efficient alternatives to \subprocessexec
that can accelerate test generation while preventing low-level crashes.

In our experiments, the baseline \pynguin achieves higher branch
coverage than \pyEvosuite in {\muTester}’s evaluation. This may be
because {\muTester} used an older \pynguin version or due to
differences in the evaluated modules---{\muTester} does not specify
them---or in constraint accuracy~(they extracted their own, we used a
set from prior work~\cite{xieDocTerDocumentationguidedFuzzing2022}).

\subsection{RQ2: Compliant Tests}

\Cref{tab:invalid_test_cases} shows the number of total, \valid, and
\invalid test cases generated by \pynguin and \pynguinml.  With
\pynguin, \PynguinValidTorch of \PynguinGenTorch test cases for
\pyTorch and \PynguinValidTf of \PynguinGenTf test cases for
\tensorflow are \valid.  With \pynguinml, \PynguinmlValidTorch of
\PynguinmlGenTorch test cases for \pyTorch and \PynguinmlValidTf of
\PynguinmlGenTf test cases for \tensorflow are \valid.  Thus,
\Pynguinml on average generates more \valid test cases in the same
amount of time than \Pynguin. Interestingly, \Pynguinml also generates
more \emph{\invalid} test cases than \pynguin: \PynguinmlInvalTorch
against \PynguinInvalTorch for \pyTorch and \PynguinmlInvalTf against
\PynguinInvalTf for \tensorflow.  This can be explained by the
generally higher code coverage achieved by \pynguinml which leads to
overall more test cases in the same time.

\begin{table}[t]
  \centering
  \caption{Generated, \valid~(\Val) and \invalid~(\InVal) test cases from
    \pynguin and \pynguinml for \exnum{10} randomly selected
  modules per ML library and the inter rater agreement as Cohen's~\cohens.}
  \label{tab:invalid_test_cases}
  \scalebox{0.98}{\begin{tabular}{@{}lrrr rrr S@{}}
  \toprule
  \textbf{Library} &
  \multicolumn{3}{c}{\textbf{Pynguin}} &
  \multicolumn{3}{c}{\textbf{PynguinML}} &
  \textbf{Cohen's~\cohens} \\
  \cmidrule(lr){2-4} \cmidrule(lr){5-7}
  & \textit{Generated} & \textit{\Val} & \textit{\InVal} &
  \textit{Generated} & \textit{\Val} & \textit{\InVal} &
  \\
  \midrule
  \pyTorch    & 154 & 107 & 42 & 180 & 121 & 45 & 0.720000 \\
  \tensorflow & 114 & 75  & 30 & 124 & 80  & 36 & 0.760000 \\
  \midrule
  \textbf{Total} & \textbf{268} & \textbf{182} & \textbf{72} &
  \textbf{304} & \textbf{201} & \textbf{81} & \textbf{0.730} \\
  \bottomrule
\end{tabular}
}
\end{table}

\begin{summary}{\hypersetup{linkcolor=white}\ref{rq:invalid}:
  \VALID Tests}
  \pynguinml generates both more \valid and more \invalid test cases in
  the same amount of time as
  \pynguin.
\end{summary}

\paragraph*{Discussion}

Our results demonstrate that using API constraints from ML library
documentation during test generation increases the number of both
\valid and \invalid test cases.
By leveraging expected parameter and tensor constraints,
\pynguinml produces more \valid inputs than \pynguin, which lets execution
pass input checks and exercise deeper code paths.

At the same time, \pynguinml deliberately produces more \invalid test cases
resulting in similar overall percentages of \valid tests for
\Pynguinml~(\pynguinmlInvalidPerc) and \Pynguin~(\pynguinInvalidPerc),
with a slight edge to \pynguin.
Our configuration intentionally allows a probability of generating
arbitrary, potentially \invalid parameters~(\qty{25}{\percent} in our
experiments).
Such \invalid test cases are beneficial because they probe input validation
logic and error-handling routines that would remain untested if using
only \valid inputs.
The additional \invalid test cases may result from the generation of
tensor-based inputs that violate constraints and from crossover and
mutation of \valid and \invalid test cases.
Overall, \valid and \invalid test cases serve complementary roles:
\Valid test cases verify that the \SUT\xspace behaves as expected,
whereas \invalid test cases evaluate its robustness.
Combined, they enable comprehensive testing of the \SUT\@.
Thus, unlike \muTester~\cite{narayananAutomaticUnitTest2023a}, we do not
aim to reduce the number of \invalid test cases.

We further observe that both \pynguin and \pynguinml
generate more test cases for \tensorflow than for \pyTorch.
This difference may reflect variations in library structure,
the number of available APIs, the strictness of internal input checks,
and the coverage and accuracy of extracted constraints.
Such differences suggest that the effectiveness of constraint-guided
test generation can depend on the characteristics of the target library.

\section{Related Work}\label{sec:related}

Previous research explored mining constraints from documentation
and using them to improve automated testing.
We used the constraints mined for the
\docTer~\cite{xieDocTerDocumentationguidedFuzzing2022} tool,
a framework for testing API functions using
fuzzing~\cite{xieDocTerDocumentationguidedFuzzing2022}. \docTer
uses the mined and pre-processed constraints to generate \valid
and \invalid test inputs for the \SUTs.
While their approach uncovered numerous bugs and inconsistencies,
fuzzing does not generate unit tests, which are necessary for
regression testing and continuous integration.

\MuTester~\cite{narayananAutomaticUnitTest2023a} overcomes this
limitation by extending the unit-test generator
\pynguin~\cite{lukasczyk_automated_2020} with API constraints,
similar to our approach, \pynguinml.
\MuTester applies linguistic rules and frequent itemset mining, achieving
improvements in code coverage by \qtyrange{15.7}{27.0}{\percent}
and reducing \invalid tests by \perc{19.0} across ML frameworks like
\tensorflow and \pyTorch.
However, the provided source code for
\MuTester\footnote{%
  \href{https://zenodo.org/records/7987893}{\url{https://zenodo.org/records/7987893}},
  last accessed 2025–08–21.%
}
is incomplete.
This makes using their framework or
replication of their experiments impossible. Nevertheless, the
results reported in their work serve as a
baseline for comparison which we used to show that \pynguinml outperforms
\MuTester in terms of code coverage.
\pynguinml is publicly available to support reproducibility~\cite{dataset}.

\section{Conclusions and Future Work}\label{sec:conclusion}

In this paper, we introduced \pynguinml, an extension of the
automated unit test generator \pynguin, designed to improve
test generation for ML libraries by leveraging API constraints
extracted from documentation.
We evaluated \pynguinml on two ML frameworks,
\pyTorch and \tensorflow, and compared it against the
baseline \pynguin and the related tool \muTester.
Our evaluation showed that \pynguinml outperforms
\Pynguin and \muTester.
\Pynguinml achieves up to \MlRelativeCoverage higher branch coverage
than \pynguin.

Although \pynguinml improves over \pynguin, many branches remain
uncovered due to the complexity of the \SUTs. Future work should
focus on efficiently generating tests for such complex APIs.
Besides ML libraries, {\pynguinml}'s flexible architecture allows
integrating constraints from other domains, like databases or web
APIs. Mining documentation to extract machine-readable constraints is
practical for well-structured libraries but might be challenging for smaller
or less-documented ones. By adapting type mappings and defining
domain-specific requirements, \pynguinml can be extended to other domains.

\section{Data Availability}\label{sec:eval:data}

We provide additional artefacts consisting of \pynguinml, datasets,
result data and analysis scripts to Zenodo to allow for future
use and replication~\cite{dataset}.

\begin{credits}
  \subsubsection{\ackname} This work was partially supported
  by the German Research Foundation (DFG)
  under grant FR 2955/5-1~(TYPES4STRINGS: Types For Strings).

  \subsubsection{\discintname}
  The authors have no competing interests to declare that are
  relevant to the content of this article.
\end{credits}

\bibliographystyle{splncs04}
\bibliography{rebibered-2}

\end{document}